\documentclass[twoside,web]{ieeecolor}
\usepackage[table,x11names,usenames,dvipsnames]{xcolor}
\usepackage[utf8x]{inputenc}
\usepackage{tmi}
\usepackage{cite}
\usepackage{amsmath,amssymb,amsfonts}
\DeclareMathOperator*{\argmax}{arg\,max}

\usepackage{subcaption}
\usepackage{tabularx}
\usepackage{algorithmic}
\usepackage{graphicx}
\usepackage{textcomp}
\usepackage{svg}
\usepackage{bm}
\usepackage{float}
\usepackage{comment}

\def\BibTeX{{\rm B\kern-.05em{\sc i\kern-.025em b}\kern-.08em
    T\kern-.1667em\lower.7ex\hbox{E}\kern-.125emX}}
\begin{document}
\title{SpecstatOR: Speckle statistics-based iOCT Segmentation Network for Ophthalmic Surgery}
\author{Kristina Mach, Hessam Roodaki, Michael Sommersperger, Nassir Navab
\thanks{This work was funded by the Bavarian Research Foundation (BFS) under grant number AZ-1569-22.}
\thanks{Kristina Mach, Michael Sommersperger, and Nassir Navab are with the Chair for Computer Aided Medical Procedures and Augmented Reality(I16), TUM School of Computation, Information and Technology, Technische
Universitat Munchen, Boltzmannstr. 3, 85748, Garching, Germany (e-mail: kristina.mach@tum.de; nassir.navab@tum.de).}
\thanks{Hessam Roodaki is with the Advanced Development department of Carl Zeiss Meditec AG, Kistlerhofstr. 75, 81379 Munich, Germany (hessam.roodaki@zeiss.com).}
}
 
\maketitle

\begin{abstract}
This paper presents an innovative approach to intraoperative Optical Coherence Tomography (iOCT) image segmentation in ophthalmic surgery, leveraging statistical analysis of speckle patterns to incorporate statistical pathology-specific prior knowledge. Our findings indicate statistically different speckle patterns within the retina and between retinal layers and surgical tools, facilitating the segmentation of previously unseen data without the necessity for manual labeling. The research involves fitting various statistical distributions to iOCT data, enabling the differentiation of different ocular structures and surgical tools. The proposed segmentation model aims to refine the statistical findings based on prior tissue understanding to leverage statistical and biological knowledge. Incorporating statistical parameters, physical analysis of light-tissue interaction, and deep learning informed by biological structures enhance segmentation accuracy, offering potential benefits to real-time applications in ophthalmic surgical procedures. The study demonstrates the adaptability and precision of using Gamma distribution parameters and the derived binary maps as sole inputs for segmentation, notably enhancing the model's inference performance on unseen data.
\end{abstract}


\section{Introduction}
\label{sec:introduction}
Optical Coherence Tomography (OCT) is a pivotal advancement in ophthalmology, offering high-resolution cross-sectional retina images. This non-invasive imaging modality has been instrumental in deepening our understanding of retinal micro structures, thereby revolutionizing diagnostic and therapeutic approaches in eye care \cite{huang1991optical}. The evolution of intraoperative OCT (iOCT) from its inception in time-domain modalities, which required mechanical adjustments for A-scan acquisition, to the adoption of more advanced spectral-domain modalities, highlights its dynamic integration within clinical settings \cite{gabriele2011optical,podoleanu2012optical}.

In the context of iOCT, enhancing images to delineate critical ocular structures distinctly is important. This enhancement facilitates the differentiation of various tissues and surgical instruments during interventions, potentially heralding advancements in fields such as autonomous robotics, signal enhancement, and advanced contextual renderings \cite{Eganathan2010RoboticTechnology,mach2022oct,He2021RoboticsSurgicalTraining}. For instance, in autonomous robotics, precise segmentation of iOCT images can enable more accurate and real-time guidance systems for surgical robots, enhancing their ability to perform delicate retinal surgeries with minimal human intervention. This can lead to improved surgical outcomes and faster recovery times for patients. 

Segmentation of iOCT acquisitions remains a challenge, whether in two or three dimensions. This is partially attributed to the inherent variability and speckle noise prevalent in these images, making it essential to explore robust statistical methods to improve accuracy. While traditional segmentation strategies range from simple threshold techniques to complex machine-learning algorithms, they frequently falter in real-time and surgical applications due to the non-rigid and mutable nature of eye structures, the diversity of surgical tools, and variability across OCT devices. The research is grounded in the premise that leveraging the intrinsic properties of light reflection and changing scattering behavior in different biological tissues can significantly aid in iOCT image segmentation. This approach harnesses the unique patterns of light interaction with tissues to create a method of segmentation that is invariant to the preservation of the shape and look of the tissue throughout a surgery, reducing the need for extensive manual intervention.

Since intraoperative applications such as navigation are the main target of this work, time-efficient methods are required to segment the iOCT in real-time. Deep learning has been increasingly utilized to address the limitations of slower algorithms while improving segmentation performance. Two notable architectures in this context are U-Net, an encoder-decoder design well-suited for biomedical image segmentation that has been adapted for real-time applications \cite{ronneberger2015unet}. ReLayNet is another fully convolutional network that goes beyond just segmenting retinal layers and also segments fluids, offering a more comprehensive approach.\cite{roy2017relaynet}

To our knowledge, no works harness the intrinsic statistics of iOCT speckles to segment ophthalmic tissue and instruments. Current literature highlights efforts to do quite the opposite and denoise diagnostic and intraoperative OCT scans from speckles \cite{mittal2022oct,koresh2021modifiedcapsule,zhou2022dhnet,nienhaus2023live}. In contrast to the prevailing focus on speckle reduction for clarity, the methodology embraces the information inherent in light-tissue interactions. 

However, there are studies analyzing and utilizing the speckles of OCT for medical purposes. A dermatology study involving OCT speckle characterization for skin tissue tried out the Rayleigh, Lognormal, Nakagami, and Generalized Gamma distributions, which determined that the Gamma distribution best characterizes skin layers.\cite{5653019}. Danilo A. Jesus et. al \cite{Jesus2015AgeRelated} determined a way to distinguish age-related structural and anatomical changes in the cornea based on speckle statistics. He used the Burr-, Gamma-, and Generalized Gamma distributions. He then did a statistical modeling of the corneal micro-structure in vivo based on OCT, which was established on Generalized Gamma distribution. \cite{Jesus2016Assessment}. Similarly, the Gamma distribution has been effectively applied to cancer images in OCT \cite{lindenmaier2013texture}.  A study from Niemczyk \cite{Niemczyk2021IntraocularPressureOCT} investigated statistical modeling of corneal diagnostic OCT to evaluate the influence of the increase in intraocular pressure and utilized the Gamma distribution. Danielewska et al. \cite{Niemczyk2021IntraocularPressureOCT}  ascertain the influence of IOP in untreated and cross-linked rabbit eyes. The Generalized Gamma distribution achieved the best goodness of fit against Rayleigh, Weibull, Nakagami, and Gamma distributions. \cite{Silva2022SignalCarrying}. Neither paper aimed at segmenting the retinal layers. One innovative approach to segment images with reduced manual work is using pseudo-labels, which allows models to learn from larger datasets without the need to label the entire dataset \cite{huang2023pseudosegrt}. This method, however, requires re-labeling and re-training for different OCT devices and does not consider the changing anatomy during surgery.

Despite the technological advancements and the increasing use of iOCT in surgical settings, challenges persist, particularly in the effective segmentation of iOCT images. Traditional methods often struggle with the inherent variability of iOCT data, leading to sub-optimal segmentation on unseen retinal structures. Recognizing this gap, our work introduces a novel approach that leverages the intrinsic statistics of iOCT speckles. Unlike conventional methods that primarily focus on speckle reduction to enhance clarity, the methodology capitalizes on the rich information the speckle patterns provide. This innovative strategy preserves and utilizes the natural properties of light-tissue interaction, enabling more accurate and real-time segmentation of tissue and surgical instruments. The approach enhances the accuracy of tissue and tool segmentation by harnessing statistical models that reflect the actual behavior of light within the eye. It paves the way for tissue-adaptive and dynamic surgical segmentation.

\section{Methods}
\label{sec:methods}
Our methodology leverages the differential scattering behaviors of light reflected from various parts of the medium to delineate distinct segmentations within a B-scan. We conduct a patch-wise distribution fitting to the speckle patterns present in the iOCT B-scan images. We analyze the amplitude and intensity histograms for each patch using maximum likelihood estimation. This robust statistical technique determines the parameters of a probability density function (PDF) by maximizing the associated likelihood or log-likelihood function. We systematically compile the estimated parameters for each patch into a 2D matrix corresponding to the B-scan dimensions.

\begin{figure}[h]  
  \centering
  % First image
    \includegraphics[width=\columnwidth]{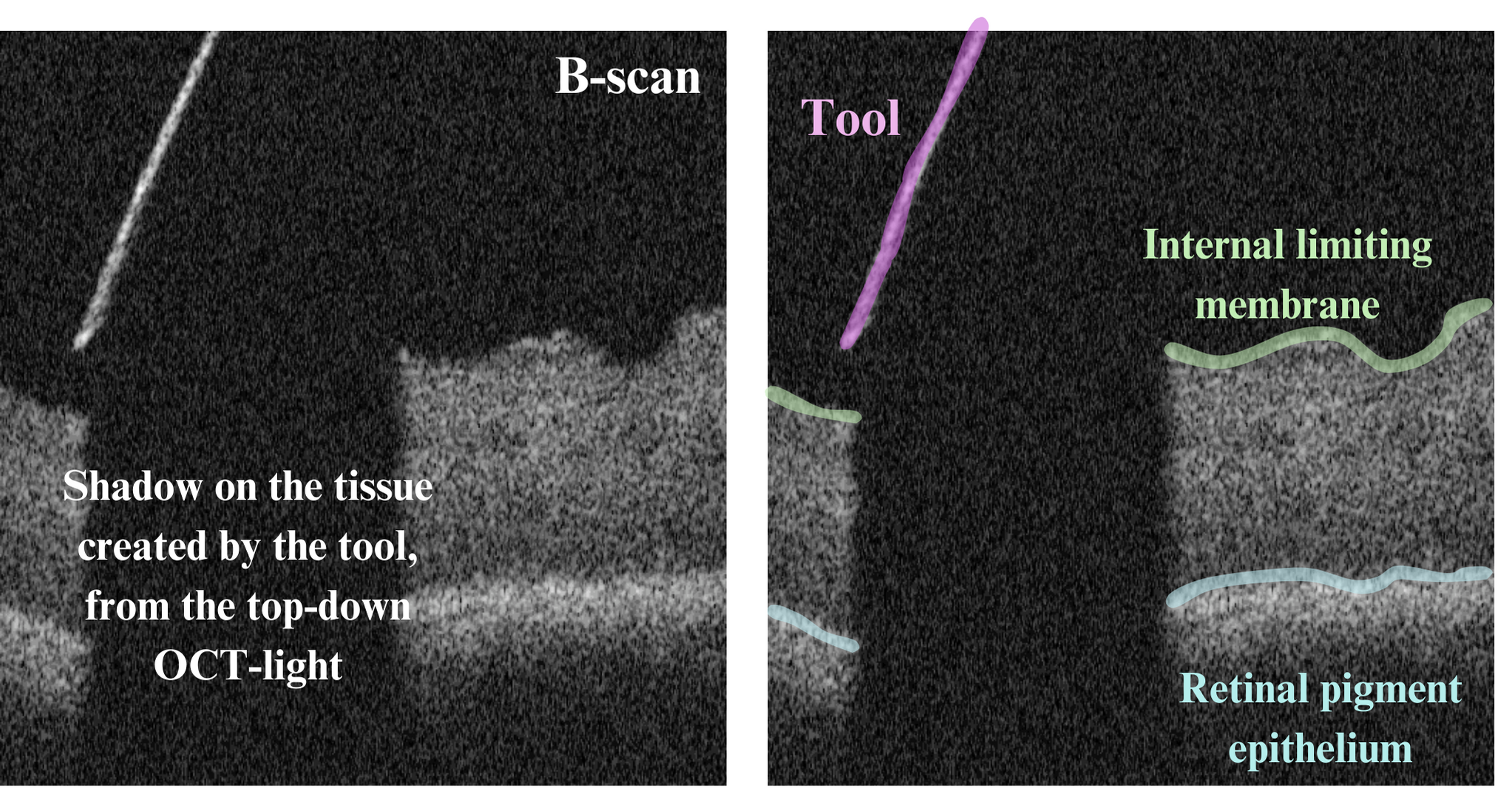} % Example image
    \caption{Retinal ex-vivo porcine iOCT B-scan with ILM (green), RPE (light blue), and Tool (Purple) classes marked.}
    \label{fig:image1}
  % Second image
    \includegraphics[width=\columnwidth]{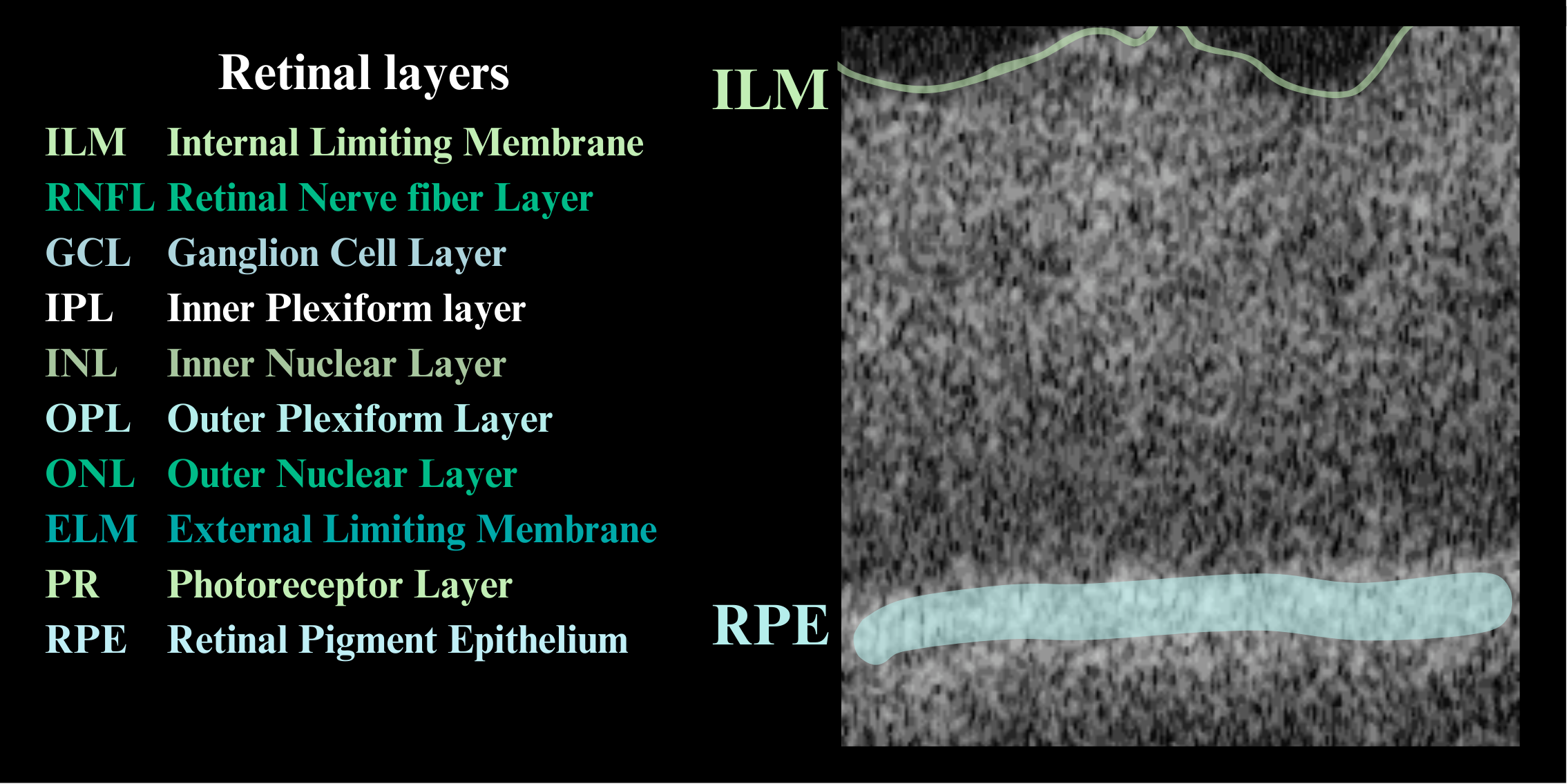} % Example image
    \caption{Ex-vivo porcine retinal layers imaged by iOCT in increasing depth from the topmost (ILM - marked in green) to the bottom most (RPE - marked in light blue).}
    \label{fig:image2}
\end{figure}

We use the Kolmogorov-Smirnov (KS) and Cram\'er–von Mises (CVM) tests to assess the accuracy of these fitted distributions, which evaluate the goodness of fit. Based on the outcomes of these tests, we categorize parameters from the best-fitting distributions into distinct value ranges. We then use these ranges to generate binary label maps, facilitating the segmentation of different anatomical and instrumental structures within the B-scan. We automate this parameter range isolation process using a random forest algorithm, classifying the parameters into predefined categories. The classes selected for segmentation are the Inner Limiting Membrane (ILM), the Retinal Pigment Epithelium (RPE), and the Tool used in the procedure. The ILM and RPE, which are the top and bottom layers of the retina, as seen in Fig. \ref{fig:image2}, act as a physical barrier when targeting intravitreal injections, the most common invasive treatment used today in ophthalmology. \cite{Irigoyen2022}

The preliminary binary maps the random forest algorithm creates serve as initial segmentations, which we subsequently refine using biological insights into retinal structures through a deep learning network. We input the 2D parameter matrices derived from the speckle statistics into another neural network. This network is designed to create a generalized model that is independent of the retinal shape, relying solely on the statistical characteristics of the speckles. This approach enhances the model's adaptability and accuracy in segmenting iOCT images across diverse surgical scenarios.

\subsection{Heterogeneity assessment}
We conduct ANOVA primarily to justify treating different classes separately in the segmentation algorithm. If ANOVA reveals significant heterogeneity between the class categories ILM, Tool, and RPE across B-scans, as shown in Fig. \ref{fig:image1}, it confirms that we can segment these classes separately due to their distinct speckle characteristics \cite{lowry2014inferential}. This step is critical to ensure the validity of subsequent goodness-of-fit analyses by establishing baseline homogeneity in pixel intensities, thereby enabling more accurate modeling of the underlying distribution of iOCT data.

Let $\bm{X}_{ijk}$ denote the pixel intensity for the $k^{th}$ observation (pixel) in the $j^{th}$ B-scan of the $i^{th}$ class, where $i \in {\text{Class 1, Class 2}}$, $j \in {1, 2, \ldots, m}$, and $m$ is the number of B-scans. The F-statistic is calculated based on the between-class variance and within-class variance as follows:

\begin{equation}
F = {\frac{MS_{\text{between classes}}}{MS_{\text{within classes}}}}.
\label{eq_anova}
\end{equation}

In \eqref{eq_anova} $MS_{\text{between classes}}$ represents the mean square for the difference between the classes, and $MS_{\text{within classes}}$ represents the mean square within each class, defined by:

\begin{equation}
MS_{\text{between classes}} = {\frac{\sum_{i} n_i (\overline{X}_i - \overline{X})^2}{m - 1}}.
\label{eq_between}
\end{equation}

\begin{equation}
MS_{\text{within classes}} = {\frac{\sum_{i} \sum_{j} \sum_{k} (X_{ijk} - \overline{X}_{ij})^2}{N - m}}.
\label{eq_within}
\end{equation}

In \eqref{eq_between} and \eqref{eq_within}, $\overline{X}i$ is the mean pixel intensity for the $i^{th}$ class, $\overline{X}{ij}$ is the mean pixel intensity for the $j^{th}$ B-scan in the $i^{th}$ class, $\overline{X}$ is the grand mean pixel intensity across all B-scans and classes, $n_i$ is the number of pixels in the $i^{th}$ class and $N$ is the total number of pixel observations across all classes and B-scans.

This methodological approach identifies significant pixel intensity variations between different anatomical structures or tools within the iOCT images. We performed the computations using the\texttt{ f$\_$oneway} function from the stats module of Python's SciPy library, enabling the calculation of the F-statistic and associated p-values for the between-class variability \cite{scipy}.

\subsection{Homogeneity Assessment}
Prior to asserting any conclusions about the speckle statistics within the ILM, RPE, and tool classes via ANOVA, we incorporated Levene's test as a fundamental step to evaluate the homogeneity of these regions across various B-scans to ensure consistency in the observations. Levene's test is essential for confirming that the variability observed in speckle patterns across different B-scans is not due to intrinsic differences in variance within a class but rather reflects genuine differences between classes \cite{levene1960robust}. By establishing that the variances within each class are equal, Levene's test safeguards against potential biases in the statistical analysis and ensures that the segmentation decisions are based on true differences in speckle patterns, not on artifacts of unequal variances among samples. 

We used the median to measure central tendency, which helps mitigate the influence of outliers, assuming the data was not normally distributed. We set the proportion of data to trim at 5\%, allowing for a conservative assessment of variance equality. We calculate the Levene's test statistic as follows:

\begin{equation}
W = {\frac{(N_i - m)}{(m - 1)} \frac{\sum_{j=1}^{m} n_{ij} (Z_{ij\cdot} - Z_{i\cdot\cdot})^2}{\sum_{j=1}^{m} \sum_{k=1}^{n_{ij}} (Z_{ijk} - Z_{ij\cdot})^2}}.
\label{eq_levenes}
\end{equation}

Equation \eqref{eq_levenes} consists of: $N_i$, which is the total number of observations in the $i^{th}$ group across all B-scans, $m$ is the number of B-scans for the $i^{th}$ group, $n_{ij}$ is the number of observations (pixel intensities) in the $j^{th}$ B-scan within the $i^{th}$ group, $Z_{ijk}$ denotes the absolute deviation of the $k^{th}$ observation from the median in the $j^{th}$ B-scan of the $i^{th}$ group, $Z_{ij\cdot}$ is the mean of the $Z_{ijk}$ values for the $j^{th}$ B-scan, and $Z_{i\cdot\cdot}$ is the overall mean of all $Z_{ijk}$ values within the $i^{th}$ group.

We computed Levene's test using the \texttt{levene} function from the stats module of Python's SciPy library. This function provides a test statistic and a corresponding p-value, with the null hypothesis that all groups have equal variances. \cite{scipy}

\subsection{Distribution Fitting}
\begin{figure}[h] 
  \centering
    \includegraphics[width=\columnwidth]{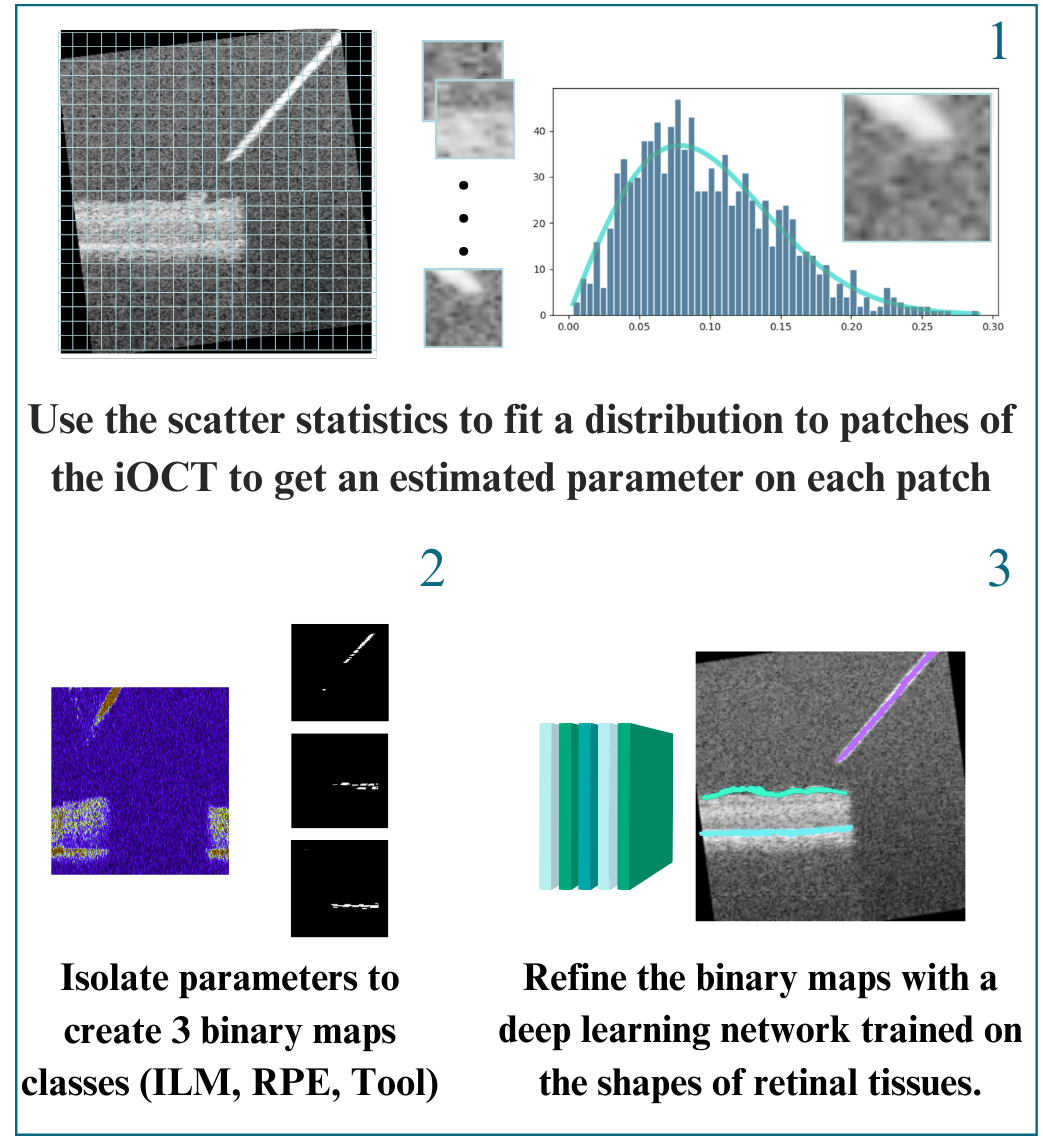} % Example image
    \caption{The pipeline of the proposed method. 1) Distribution fitting. 2) Parameter isolation to classes. 3) Refinement with deep neural network.}
    \label{fig:pipeline}
\end{figure}

We modeled the empirical distribution of pixel intensities within iOCT images to identify suitable statistical distributions for ILM, RPE, and surgical tools. To achieve this, we fitted a series of continuous probability distributions to the data, including Gamma, Rayleigh, Normal, Burr, Lognormal, and Nakagami distributions. We show this first step in our pipeline in Fig. \ref{fig:pipeline}. We selected these distributions based on previous literature in diagnostic OCT and their relevance to the light scattering and absorption properties of ocular tissues and surgical instruments observed in iOCT images. \cite{Niemczyk2021IntraocularPressureOCT,grzywacz2010statistics, lindenmaier2013texture}.

We executed the parameter fitting on 7x7-sized patches on the B-scans. We modeled each patch region of scatters using a histogram and fitted it using Maximum Likelihood Estimation (MLE), which seeks to find values that maximize the likelihood function given the observed data. This process resulted in a new matrix for each distribution parameter or parameters, each consisting of 7x7 patches optimized to the value of the distribution. The MLE approach is beneficial as it provides estimators with good asymptotic properties, such as consistency and asymptotic normality, which are desirable for large sample sizes like those in iOCT image data sets.

\begin{equation}
\hat{\theta}_{\text{dist}} = {\argmax_\theta L(\theta | x)}.
\label{eq_mle}
\end{equation}

In \eqref{eq_mle}, $\hat{\theta}_{\text{dist}}$ represents the estimated parameters for the given distribution 'dist,' $L(\theta | x)$ is the likelihood of the parameters given the observed pixel intensities $x$, and $\argmax_\theta$ denotes the argument of the parameters that maximize the likelihood function.

We estimated the shape ($k$) and scale ($\theta$) parameters for the Gamma distribution. The Normal distribution required fitting the $mean$ and $standard deviation$. For the Burr distribution, they fitted the two parameters ($c$ and $d$), resulting in two matrices per distribution. They also estimated the scale parameter ($\sigma$) for the Rayleigh distribution, the parameter ($s$) for the Lognormal distribution, and the parameter ($\nu$) for the Nakagami distribution, each producing one matrix per distribution.

\subsection{Goodness-of-Fit Analysis}
Following the distribution fitting, we performed a goodness-of-fit analysis using the Kolmogorov-Smirnov (KS) statistic and the Cram\'er–von Mises criterion. To measure the maximum discrepancy between the empirical distribution function (EDF) of the sample and the cumulative distribution function (CDF) of the hypothesized statistical model, we applied the Kolmogorov-Smirnov (KS) test. This test is especially useful in this context because of its non-parametric nature, offering a sensitive fit measure without assuming the distribution's parameters \cite{Kolmogorov}. The Kolmogorov-Smirnov statistic is defined as:

\begin{equation}
KS = {\sup_x |F_n(x) - F(x)|}.
\label{eq_KS}
\end{equation}

In \eqref{eq_KS}, $KS$ is the KS statistic representing the supremum of absolute differences between the EDF ($F_n(x)$) of the sample and the CDF ($F(x)$) of the assumed model.

The Cram\'er–von Mises criterion, on the other hand, assesses the sum of squared discrepancies across the distribution's entire range, providing a rigorous evaluation of the overall fit between the empirical data and the theoretical model \cite{anderson1962distribution}. 

\begin{equation}
CVM = {\frac{1}{12N} + \sum_{i=1}^{N} \left[\frac{2i - 1}{2N} - F\left(x_{(i)}\right)\right]^2}.
\label{eq_cramer}
\end{equation}

In \eqref{eq_cramer}, $CVM$ is the Cram\'er–von Mises statistic, $N$ is the number of observations within the group, and $x_{(i)}$ are the ordered observations. This statistic provides a global assessment of how well the hypothesized distribution models the entire data set.

To ensure a comprehensive approach, we selected these two goodness-of-fit tests: the KS test focuses on the most extreme deviations, sensitive to central distribution differences, while the Cram\'er–von Mises test checks for discrepancies across the entire distribution, enhancing sensitivity to variations in the tails. We calculated the tests using functions from Python's SciPy library, comparing each group's pixel intensity distributions against their respective theoretical models \cite{scipy}. A high p-value from these tests indicates a good fit, suggesting the hypothesized distribution appropriately models the pixel intensity data for that group.

\subsection{Classification}
Distribution fitting assumes that sufficiently different parameters for ILM, RPE, and Tool classes will emerge. Thus, unique parameter combinations for each class can be identified and isolated into three different binary maps, as illustrated in the pipeline's second step in Fig. \ref{fig:pipeline}. Although manual isolation of parameters is possible, the proposed method employs a Random Forest algorithm to enhance time efficiency. Five volumes were randomly selected for generalization, and 10 B-scans per volume were trained using the Random Forest algorithm. Inputs to the Random Forest algorithm included the 2D patch-wise parameter matrices, with ground truth based on corresponding patch-wise ground truth segmentations. Training established logical rules that differentiated parameter ranges best capturing the three classes—ILM, RPE, and Tool—thereby creating segmentation maps. The algorithm was subsequently applied to all remaining volumes, generating what are referred to as weak labels.

\begin{figure}[h]
        \includegraphics[width=\columnwidth]{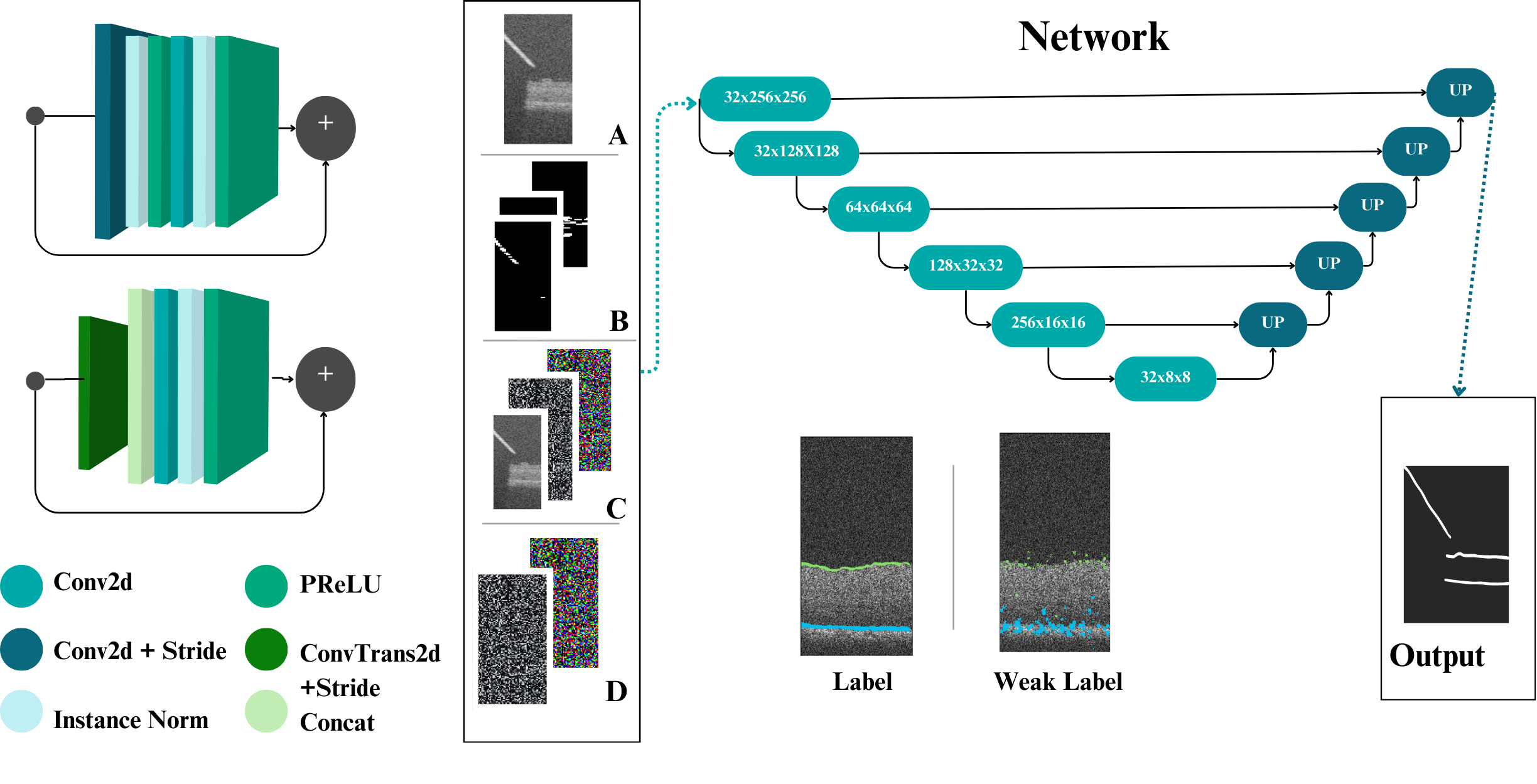}
        \caption{Network architecture and training configurations A, B, C, and D as different inputs. Labels: weak from the parameter isolation and segmented ground truth.}
    \label{fig:networkarchitecture}
\end{figure}

Though initially coarse, these weak labels undergo refinement through a deep-learning network. This step integrates insights from statistical and physics-based analyses with a biological understanding of the retina, depicted as step 3 in Fig. \ref{fig:pipeline}. Employing an enhanced version of U-Net, which includes residual units known for their effectiveness in biomedical image segmentation, allowed for precise refinement \cite{kerfoot2019leftventricle}. The architecture of U-Net is specifically designed to produce three binary maps representing ILM, RPE, and the surgical tool, critical for refining segmentation outcomes based on goodness-of-fit analysis parameters, as shown in Fig. \ref{fig:networkarchitecture}.

First model variation:
\begin{itemize}
    \item \textit{B:} Label maps created from a statistical understanding of the OCT.
    \item \textit{Output:} Perfected label maps of ILM, RPE, and Tool based on a biological understanding of the retina.
\end{itemize}

The analysis also covered whether weak labels could serve in segmentation models where no ground truth data is available, thus potentially automating the creation of labeled data for training large models. This involved calculating the model's loss function based on discrepancies between the output and the weak label created from the Random Forest algorithm's parameter isolation. An ablation study assessed the impact of various input data configurations, as depicted in Fig. \ref{fig:networkarchitecture}. The aim was to determine whether distribution parameters alone suffice for segmentation or if B-scan information is necessary. The network, configured to output label maps of ILM, RPE, and Tool, was trained with four distinct input setups:

\begin{itemize}
    \item \textit{A:} B-scans alone as a baseline.
    \item \textit{B:} Label maps created from a statistical understanding of the OCT.
    \item \textit{C:} A combination of B-scans and Gamma distribution parameters. 
    \item \textit{D:} Gamma distribution parameters alone.
\end{itemize}

The architecture of the network can be seen in Fig. \ref{fig:networkarchitecture}. The network has 4.0 million trainable parameters, resulting in an estimated model size of 16.094 MB. The training variants can also be seen in Fig. \ref{fig:networkarchitecture}. 

\subsection{Data Acquisition and Preprocessing}
Three datasets collected using distinct OCT devices facilitate a comparative analysis. The first set, referred to as Group One, comprises 50 iOCT acquisitions obtained using a Zeiss Lumera 700 with an integrated Rescan 700 OCT system (Carl Zeiss Meditec AG). These iOCT scans, which include a volumetric collection of B-scans, each with a depth of 2.8mm, were acquired from ex-vivo porcine eyes to simulate a broad field-of-view environment. This dataset features two different surgical tools: one metallic needle and one forceps. Porcine and human eyes share many anatomic similarities, making them valuable in ophthalmic research \cite{Prince1960, Beauchemin1974}.

The second dataset, which we refer to as Group Two, is collected using Leica Proveo 8 With EnFocus OCT Imaging System (Leica Microsystems GmbH). It consists of 311 B-scans, which are comprised of 1000 A-scans with an axial resolution of 1024 pixels, each with a scanning depth of 3.38 mm, containing 100 linearly acquired B-scans. Two different metallic needles are present in the dataset.

Likewise, Group Three, utilizing the same Zeiss system as Group One, involved five different porcine eyes for validation purposes and include a glass needle in the tool class. Both Group One and Group Two include B-scans with corresponding label maps of the ILM, RPE, and tools as presented during each surgical phase. 

Groups One and Two include B-scans with corresponding label maps for the ILM, RPE, and the tools, if present, during that surgical phase. Figure \ref{fig:image1} provides a visualization of a B-scan of a porcine eye with markings for ILM and RPE. In contrast, the third dataset contains only a few labeled B-scans per volume.
 
\section{Results}
\label{sec:results}

\begin{table}
\centering
\begin{tabular}{lccl}
\hline
\textbf{Group} & \textbf{Statistics} & \textbf{P-value} & \textbf{Conclusion}                \\ \hline
\multicolumn{4}{c}{\textbf{Group One}}                                                         \\ \hline
Tool           & 1.916               & 0.046            & Variances not equal (reject H0)    \\ 
ILM            & 1.217               & 0.279            & Variances equal (fail to reject H0) \\ 
RPE            & 1.379               & 0.192            & Variances equal (fail to reject H0) \\ \hline
\multicolumn{4}{c}{\textbf{Group Two}}                                                         \\ \hline
Tool           & 8.060               & 0.000            & Variances not equal (reject H0)    \\ 
ILM            & 2.394               & 0.531            & Variances equal (fail to reject H0)   \\ 
RPE            & 1.743               & 0.074            & Variances equal (fail to reject H0) \\ \hline
\end{tabular}
\caption{Levene's Test Results for Group One and Group Two}
\label{tab:levenes_test}
\end{table}

\subsection{Levene's test}
The analysis utilizing Levene's test to assess the homogeneity of variances across different classes ILM, RPE, and Tool within two distinct iOCT B-scan Groups One and Two is presented in Table~\ref{tab:levenes_test}. Our observations delineate that variances within the B-scan groups for ILM and RPE are consistent, thus endorsing the application of ANOVA. However, noteworthy to address is a small imbalance in variance within Group Two for RPE was noted. Significant variance disparities across the B-scans for the Tool in both iOCT B-scan groups necessitate a cautious application of ANOVA. This variability was anticipated, given the tools' heterogeneous material composition and varying thicknesses, which induce non-uniform speckle patterns.

\subsection{ANOVA}
The ANOVA conducted to study the variability between and within ILM, RPE, and Tool classes across the two B-scan groups has demonstrated key insights, as detailed in Table~\ref{tab:anova_results}. In Group One, the comparison between Tool and ILM manifested a considerable difference in speckle characteristics, underscored by a notable F-value of 13.421 and a P-value of 0.00025, suggesting pronounced disparities. Conversely, the RPE versus ILM comparison indicates similar character in speckle patterns amongst these classes.

\begin{table}[h]
\centering
    \caption{Statistical ANOVA Analysis of Pixel Distributions in B-scan Regions}
\begin{tabular}{cccc}
\hline
B-scan Group & F-value & P-value & Class \\
\hline
1 & 13.42105475 & 0.00024889 & Needle ILM \\
1 & 6.15219291 & 0.01312585 & Needle RPE \\
1 & 1.86339097 & 0.17223564 & RPE ILM \\
2 & 2387.30972878 & $1.51489985 \times 10^{-22}$ & Needle ILM \\
2 & 119.01615676 & $1.06381005 \times 10^{-27}$ & Needle RPE \\
2 & 2043.63004008 & $1.42481005 \times 10^{-26}$ & RPE ILM \\
\hline
\end{tabular}
\caption{ANOVA results comparing the mean speckle intensities within each B-scan group.}
\label{tab:anova_results}
\end{table}

In Group Two, a stark contrast was observed in all pairwise class comparisons as seen in Table~\ref{tab:anova_results}, demonstrating significant statistical variances. These findings substantiate the premise of distinctive speckle characteristics between the classes, especially pronounced in Group Two. The statistical significance of these disparities validates the hypothesis that applying distinct distribution can form a segmentation by effectively characterizing speckle patterns.

\subsection{Distribution fit for Group One}

\subsubsection{ILM}
Analysis (referenced in Table~\ref{tab:CVMgroup1} and Table~\ref{tab:KSgroup1}) revealed the Burr distribution as exhibiting the highest mean p-value with minimal variance in both the Cram\'er–von Mises and Kolmogorov-Smirnov tests, indicating a consistent fit. The Gamma distribution displayed a moderate mean p-value, albeit with variance suggesting moderate inconsistency across B-scans. The Rayleigh distribution, characterized by significantly low p-values, demonstrated an inadequate model fit. The Normal distribution presented an average model fit, tempered by its variability, while the Lognormal and Nakagami distributions displayed moderate fit quality with relatively lower variability, suggesting a degree of consistency.

\begin{table}[H]
\centering
\tiny
\begin{tabular}{lllllllll}
\hline
\textbf{Type} & \textbf{Gamma} & \textbf{G P-val} & \textbf{Rayleigh} & \textbf{R P-val} & \textbf{Norm} & \textbf{N P-val} & \textbf{Class} \\ \hline
mean & 6.54E-02 & 1.38E-02 & 1.95E-01 & 5.28E-16 & 5.58E-02 & 3.34E-02 & ILM \\
variance & 1.60E-04 & 3.85E-03 & 2.77E-04 & 2.33E-29 & 1.38E-04 & 1.10E-02 & ILM \\

mean& 8.35E-02 & 1.08E-02 & 2.58E-01 & 1.29E-14 & 7.18E-02 & 2.20E-02 & RPE \\
variance& 3.33E-04 & 3.81E-03 & 1.93E-03 & 1.40E-26 & 2.57E-04 & 1.10E-02 & RPE \\

mean& 7.52E-02 & 4.05E-02 & 2.13E-01 & 8.42E-03 & 6.48E-02 & 4.05E-02 & Tool \\

variance& 2.67E-04 & 1.75E-02 & 3.15E-03 & 2.53E-03 & 1.85E-04 & 1.62E-02 & Tool \\ 

\hline
 & \textbf{Burr} & \textbf{B P-val} & \textbf{Lognorm} & \textbf{L P-val} & \textbf{Nakagami} & \textbf{N P-val} &  \\ \hline
mean & 2.97E-02 & 3.96E-01 & 5.68E-02 & 2.81E-02 & 5.58E-02 & 3.33E-02 & ILM \\
variance & 5.54E-05 & 6.06E-02 & 1.38E-04 & 7.82E-03 & 1.38E-04 & 1.10E-02 & ILM \\

mean& 3.39E-02 & 2.87E-01 & 7.24E-02 & 2.10E-02 & 7.18E-02 & 2.17E-02 & RPE \\
variance& 7.77E-05 & 5.25E-02 & 2.54E-04 & 1.03E-02 & 2.57E-04 & 1.05E-02 & RPE \\

mean& 2.99E-02 & 3.28E-01 & 6.47E-02 & 5.06E-02 & 6.44E-02 & 5.06E-02 & Tool \\ 

variance& 1.06E-04 & 5.71E-02 & 1.91E-04 & 2.44E-02 & 1.92E-04 & 2.26E-02 & Tool \\ \hline
\end{tabular}
\caption{Cram\'er-von-Mises-Test results for distributions in Group One}
\label{tab:CVMgroup1}
\end{table}

\begin{table}[H]
\centering
\tiny
\begin{tabular}{lllllllll}
\hline
\textbf{Type} & \textbf{Gamma} & \textbf{G P-val} & \textbf{Rayleigh} & \textbf{R P-val} & \textbf{Norm} & \textbf{N P-val} & \textbf{Class} \\ \hline
mean & 1.12E+00 & 2.11E-02 & 1.45E+01 & 2.29E-09 & 7.47E-01 & 5.36E-02 & ILM \\
variance & 2.65E-01 & 5.02E-03 & 2.41E+01 & 5.87E-18 & 1.28E-01 & 1.58E-02 & ILM \\

mean& 2.19E+00 & 1.55E-02 & 2.72E+01 & 5.77E-09 & 1.48E+00 & 2.81E-02 & RPE \\
variance& 1.38E+00 & 6.27E-03 & 1.68E+02 & 3.09E-17 & 6.35E-01 & 1.44E-02 & RPE \\

mean& 2.33E+00 & 4.28E-02 & 2.54E+01 & 7.58E-03 & 1.55E+00 & 5.04E-02 & Tool \\
variance& 2.19E+00 & 1.85E-02 & 2.38E+02 & 2.32E-03 & 9.85E-01 & 1.95E-02 & Tool \\ 
\hline
 & \textbf{Burr} & \textbf{B P-val} & \textbf{Lognorm} & \textbf{L P-val} & \textbf{Nakagami} & \textbf{N P-val} & \textbf{Class}\\
 \hline
mean & 1.23E-01 & 5.78E-01 & 7.76E-01 & 4.62E-02 & 7.48E-01 & 5.33E-02 & ILM \\
variance & 8.43E-03 & 5.38E-02 & 1.31E-01 & 1.14E-02 & 1.30E-01 & 1.56E-02 & ILM \\

mean& 1.76E-01 & 4.47E-01 & 1.51E+00 & 2.64E-02 & 1.48E+00 & 2.78E-02 & RPE \\
variance& 1.97E-02 & 6.14E-02 & 6.36E-01 & 1.32E-02 & 6.25E-01 & 1.39E-02 & RPE \\

mean& 1.47E-01 & 4.87E-01 & 1.57E+00 & 5.44E-02 & 1.55E+00 & 6.01E-02 & Tool \\
variance& 7.40E-03 & 5.96E-02 & 9.98E-01 & 2.39E-02 & 1.01E+00 & 2.84E-02 & Tool \\ \hline
\end{tabular}
\caption{Kolmogorov–Smirnov-Test results for distributions in B-scan Group One.}
\label{tab:KSgroup1}
\end{table}

\subsubsection{RPE}
For the RPE class, the Burr distribution's performance was notably good based on the CVM test mean p-value, yet accompanied by increased variance, hinting at fit inconsistency. The KS test for the Norm distribution suggested a modest degree of fit variability. Similarly, the Gamma and Lognormal distributions indicated a generally acceptable fit, with the Gamma distribution showing slight superiority in mean p-value compared to the ILM class despite high variance indicating fit inconsistency across samples.

\subsubsection{Tool}
The Tool class revealed a high mean p-value for the Burr distribution in the CVM test, indicating a potentially good fit, but with significant variance pointing to inconsistency. The Rayleigh distribution's consistently low p-values across B-scans underlined its unsuitability. The Normal and Lognormal distributions suggested moderate fits with substantial variance, whereas the Gamma and Nakagami distributions struggled with low mean p-values and high variability, indicating poor fits.

\subsection{Distribution fit for Group Two}

\subsubsection{ILM}
The analysis for Group Two as can be seen in (Table \ref{tab:CVMgroup2}) and Table \ref{tab:KSgroup2} mirrored the findings from Group One, with the Gamma distribution showing a good fit for the ILM class, albeit with limitations indicated by its p-value. The Rayleigh distribution, maintaining its trend, exhibited low p-values. The Normal distribution's variability suggested a conditional fit, whereas the Burr and Lognormal distributions indicated moderate fits with noticeable variability. The Nakagami distribution, similar to its performance in Group One, showed a high p-value but with significant variability.

\begin{table}[H]
\centering
\tiny
\begin{tabular}{lllllllll}
\hline
\textbf{Type} & \textbf{Gamma} & \textbf{G P-val} & \textbf{Rayleigh} & \textbf{R P-val} & \textbf{Norm} & \textbf{N P-val} & \textbf{Class} \\
\hline
mean & 6.17E+00 & 7.63E-02 & 6.77E+00 & 1.47E-03 & 1.68E+00 & 9.22E-03 & ILM \\
variance & 5.36E+02 & 8.69E-03 & 1.55E+01 & 1.36E-04 & 7.01E-01 & 7.39E-04 & ILM \\

mean& 3.06E-01 & 3.19E-01 & 1.58E+01 & 2.62E-09 & 3.62E-01 & 2.60E-01 & RPE \\
variance& 1.38E+00 & 6.27E-03 & 1.68E+02 & 3.09E-17 & 6.35E-01 & 1.44E-02 & RPE \\

mean& 2.68E+00 & 1.07E-01 & 9.90E+00 & 3.02E-03 & 6.46E-01 & 9.67E-02 & Tool \\
variance& 2.19E+00 & 1.85E-02 & 2.38E+02 & 2.32E-03 & 9.85E-01 & 1.95E-02 & Tool \\

\hline
& \textbf{Burr} & \textbf{B P-val} & \textbf{Lognorm} & \textbf{L P-val} & \textbf{Nakagami} & \textbf{N P-val} &  \\
mean & 7.03E+01 & 2.36E-03 & 5.14E-01 & 8.62E-02 & 7.03E-01 & 5.61E-02 & ILM \\
variance & 2.77E+03 & 1.49E-04 & 6.53E-02 & 1.01E-02 & 1.32E-01 & 1.13E-02 & ILM \\

mean& 6.98E+01 & 3.66E-02 & 4.58E-01 & 2.09E-01 & 2.82E-01 & 3.22E-01 & RPE \\
variance& 1.97E-02 & 6.14E-02 & 6.36E-01 & 1.32E-02 & 6.25E-01 & 1.39E-02  &RPE \\

mean& 5.92E+01 & 2.39E-02 & 1.20E+00 & 7.90E-02 & 5.85E-01 & 1.18E-01 & Tool \\
variance& 7.40E-03 & 5.96E-02 & 9.98E-01 & 2.39E-02 & 1.01E+00 & 2.84E-02 & Tool \\ \hline
\end{tabular}
\caption{Cram\'er-von-Mises-Test results for distributions in B-scan Group Two.}
\label{tab:CVMgroup2}
\end{table}

\begin{table}[H]
\centering
\tiny
\begin{tabular}{llllllll}
\hline
\textbf{Type} & \textbf{Gamma} & \textbf{G P-val} & \textbf{Rayleigh} & \textbf{R P-val} & \textbf{Norm} & \textbf{N P-val} & \textbf{Class} \\ \hline

mean & 7.11E-2 & 4.30E-2 & 1.27e-1 & 3.09e-4 & 7.02E-2 & 4.306E-3 & ILM \\
variance & 1.41E-2 & 3.99E-3 & 1.01E-3 & 4.36E-6 & 2.39E-4 & 5.03E-4 & ILM \\
mean & 3.35E-2 & 2.37E-1 & 1.81E-1 & 1.11E-12 & 3.64E-2 & 1.84E-1 & RPE \\
variance & 1.17E-4 & 6.99E-2 & 1.65E-3 & 2.63E-23 & 1.23E-4 & 5.43E-2 & RPE \\
mean & 6.10E-2 & 9.22E-2 & 1.49E-1 & 2.85E-3 & 5.15E-2 & 8.42E-2 & Tool \\
variance & 5.00E-3 & 2.52E-2 & 3.82E-3 & 4.10E-4 & 2.35E-4 & 2.OOE-2 & Tool \\ \hline
 & {\textbf{Burr}} & {\textbf{B P-val}} & {\textbf{Lognorm}} & {\textbf{L P-val}} & {\textbf{Nakagami}} & {\textbf{N P-val}} & \\ \hline
mean & 3.89e-1 & 9.42e-4 & 4.177E-2 & 4.83E-2 & 4.93E-2 & 3.19E-2 & ILM \\
variance & 3.89E-2 & 2.89e-5 & 7.68e-5 & 4.06e-3 & 1.41e-4 & 7.15e-3 & ILM \\
mean & 3.89e-1 & 3.11E-2 & 3.91E-2 & 1.55e-1 & 3.31E-2 & 2.31e-1 & RPE \\
variance & 4.20E-2 & 1.81E-2 & 1.97e-4 & 4.61E-2 & 1.01e-4 & 5.37E-2 & RPE \\
mean & 4.198e-1 & 2.42E-2 & 6.07E-2  & 6.91E-2 & 4.82E-2 & 10.46E-2 & Tool \\
variance & 4.62E-2 & 1.62E-2 & 0.70E-2 & 1.46E-2 & 0.21E-2 & 2.69E-2 & Tool \\ \hline
\end{tabular}
\caption{Kolmogorov–Smirnov-Test results for distributions in B-scan Group Two .}
\label{tab:KSgroup2}
\end{table}

\subsubsection{RPE}
For the RPE class in Group Two, the Gamma distribution's higher p-value than the ILM class suggested a better fit. The Rayleigh distribution's low p-values indicated its low fit. The Normal distribution mirrored the moderate fit observed in the ILM class, while the Burr and Lognormal distributions demonstrated moderate fits with some inconsistency, similar to their Group One performance. The Nakagami distribution's performance was marked by high variability.

\subsubsection{Tool}
The Tool class in Group Two exhibited patterns akin to those observed in Group One, with the Gamma distribution showing a low mean p-value, underscoring a sub-optimal fit. The Rayleigh distribution's low p-values reiterated its status. The Normal distribution presented a moderate mean p-value with significant variance, suggesting an inconsistent fit. The Burr and Lognormal distributions, despite showing potential in certain cases, were characterized by considerable variance, indicating inconsistency. The Nakagami distribution showed promise in some instances but was hindered by high variability.

\subsection{Class Evaluation across B-scan groups}
This refined analysis underscores the nuanced performance of statistical distributions in modeling iOCT B-scan data across two distinct groups and three classes (ILM, RPE, and Tool). The Burr distribution consistently shows promise, albeit with noted inconsistency, while the Rayleigh distribution's fit across all classes highlights its limitations. The Gamma distribution emerges as a generally favorable model, especially for the RPE class in Group Two, albeit with noted variance for the Tool class.

\subsection{Parameter extraction}
\begin{figure}[h] 
  \centering
  \includegraphics[width=8.8cm]{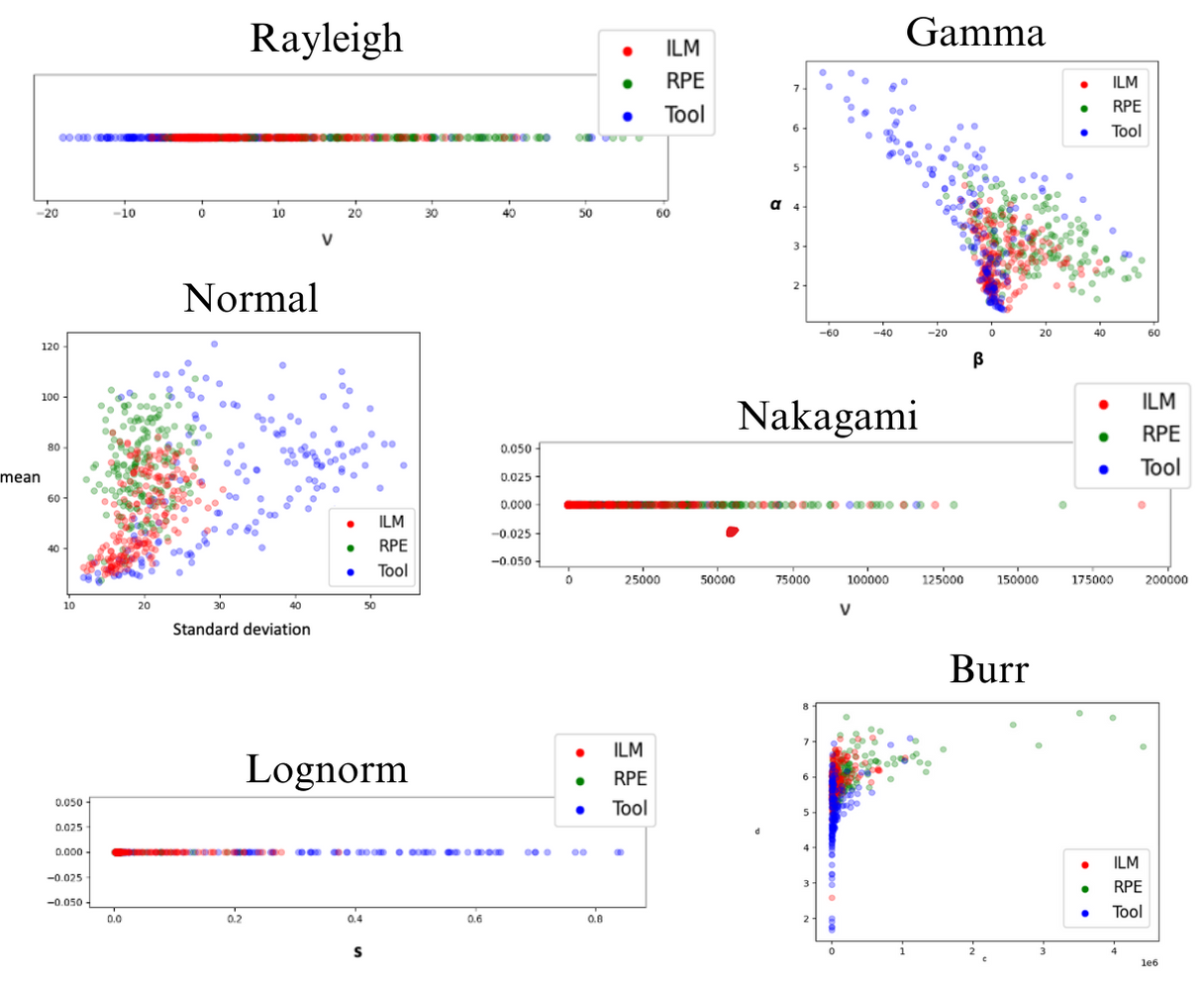}
  \caption{Scatterplots of statistical distribution parameters, fitted for Group Two}
  \label{scatterplots_distributions}
\end{figure}

The scatter plots of the fitted parameters for each distribution can be seen in Fig. \ref{scatterplots_distributions}. Nakagami, Rayleigh, and Lognormal all have one parameter to tune where the Rayleigh distribution parameters can be linearly separated into different classes. For the 2 parameter distributions, Normal, Burr, and Gamma distribution, the two more straightforward parameter separations seem to be Gamma and Normal distribution parameters.

\begin{figure}[h] 
  \centering
  \includegraphics[width=8.8cm]{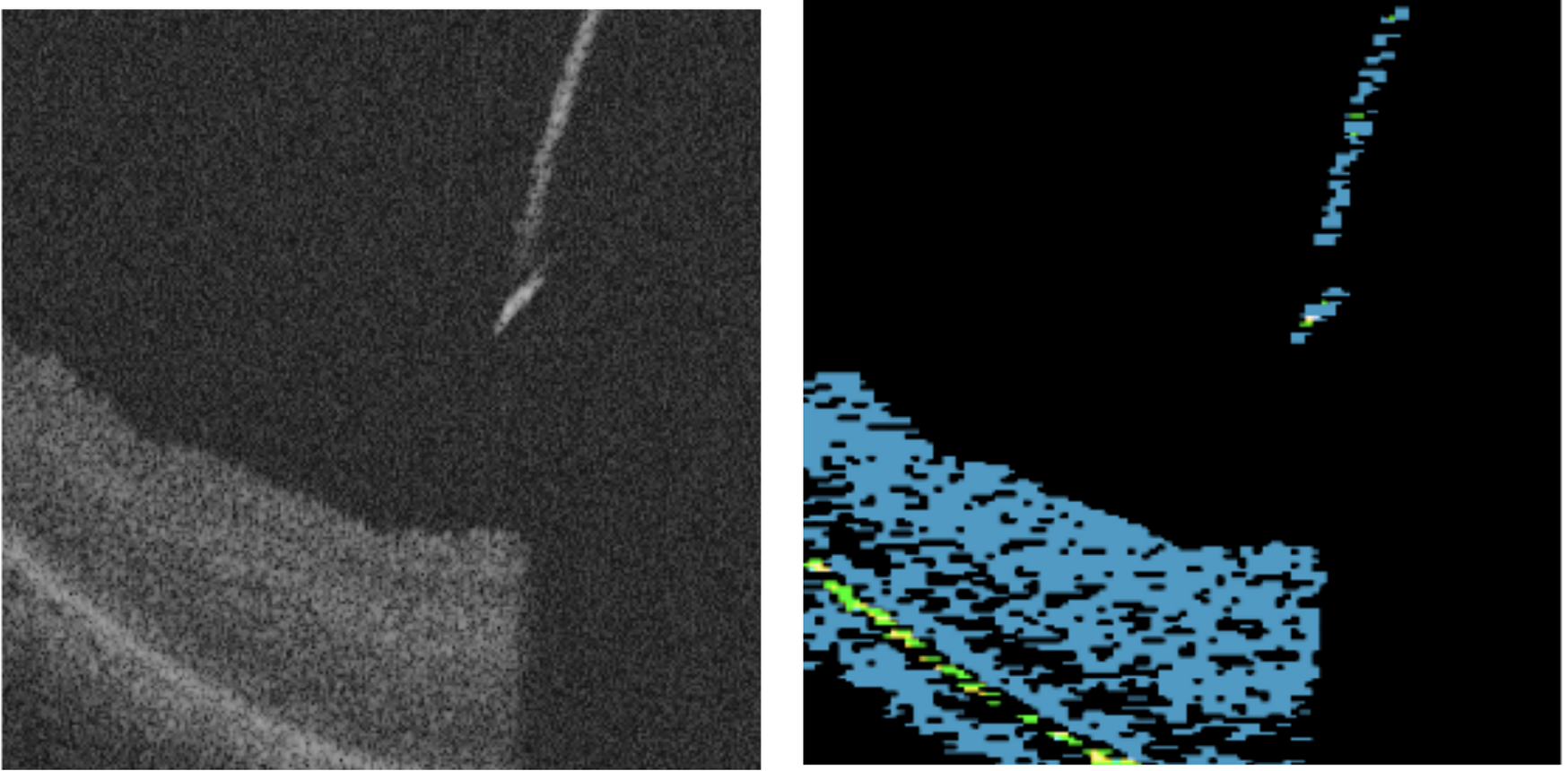}
  \caption{Calibrated Gamma distribution parameters, Group One iOCT, alpha parameters between 10 and 30 shown with green, below 10 black, and the rest blue.}
  \label{fig:segmentation_rpe_layer}
\end{figure}

\begin{figure}[h] 
  \centering
  \includegraphics[width=\columnwidth]{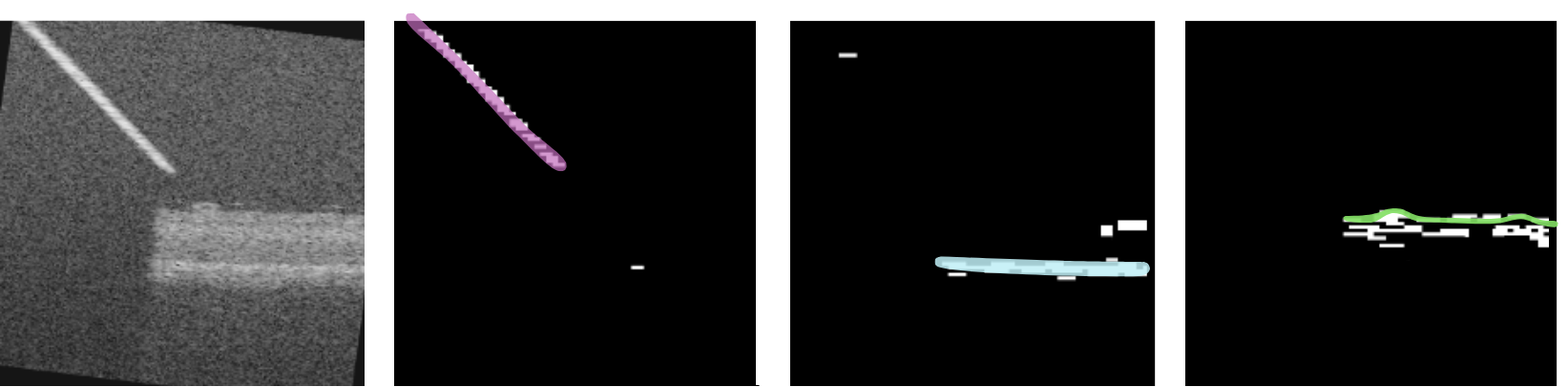}
  \caption{Calibrated Gamma distribution parameters, Group Two iOCT, alpha and beta parameters isolated into 3 different binary maps that correspond to the classes ILM (green), RPE (light blue), and Tool (purple).}
  \label{fig:segmentation_all_layers}
\end{figure}

An example of this kind of classification of specific parameter values can be seen in Fig. \ref{fig:segmentation_rpe_layer}, where we delineated the alpha parameter within the range of ten to thirty, corresponding to the RPE layers segmented and highlighted in green. Furthermore, Fig. \ref{fig:segmentation_all_layers} demonstrates how the parameter value classification can yield three distinct binary maps. Notably, due to some values being wrongly classified because of overlap, the resultant segmentation maps exhibit a degree of noise. This nuanced approach underscores the inherent variability within the data and highlights the need for meticulous parameter selection to enhance the segmentation outcome.

\subsection{Validation}
\begin{figure}[h]
\centering
    % Group One
    \includegraphics[width=\columnwidth]{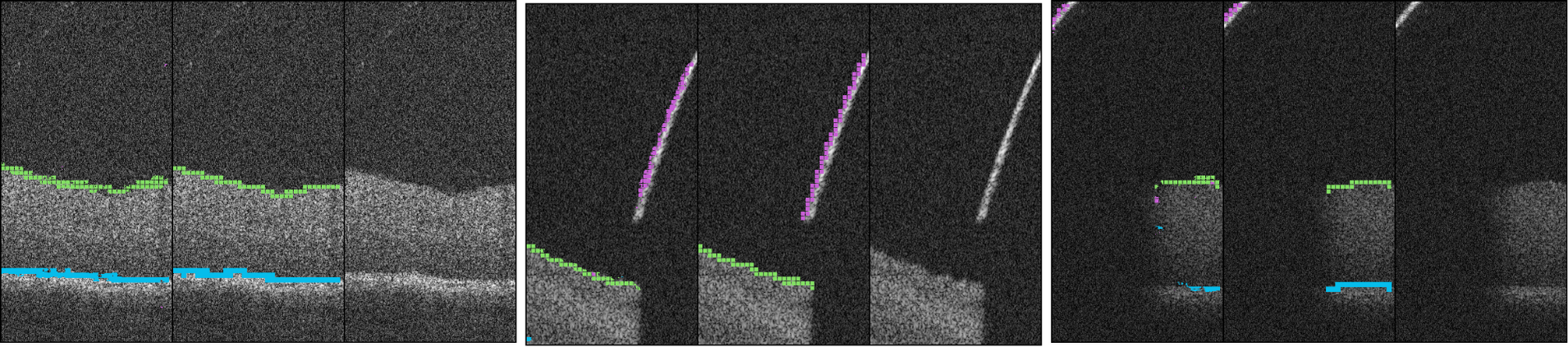}
    \includegraphics[width=\columnwidth]{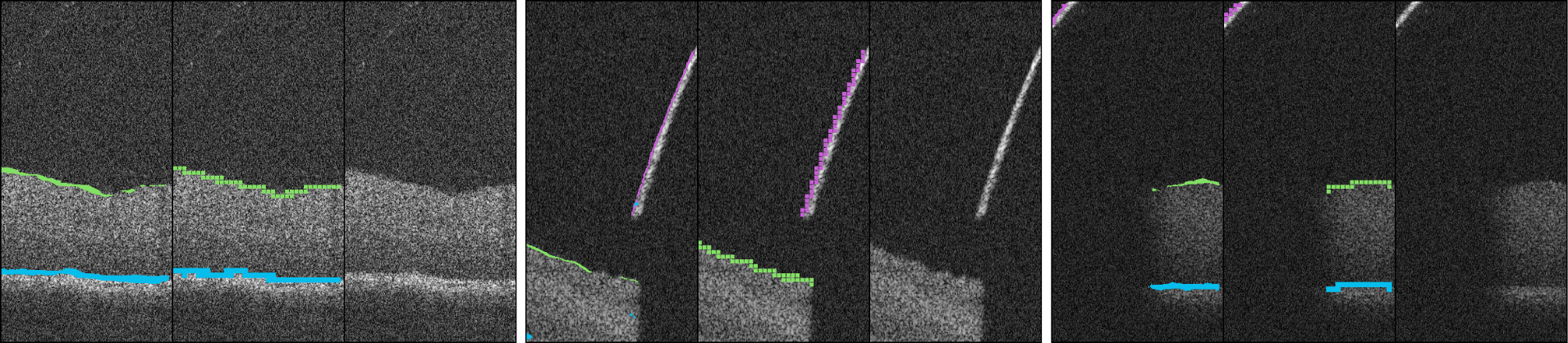}
    \caption{Three examples of Group One images with training configuration D (top) and training configuration A (bottom). In each example, the left image is the output, the middle image is the ground truth, and the right image is the raw B-scan.}
    \label{fig:group1visual}
\end{figure}

The comparison between Groups One and Two (used for decision tree development for binary map generation) and Group Three (completely new, unseen data) highlights the method's robustness. While Groups One and Two show high Dice scores and lower Hausdorff distances for all configurations, as can be seen in Fig. \ref{tab:group1and2_result}, Group Three demonstrates a broad performance spectrum, emphasizing the challenges of segmenting new datasets. Three visual examples of the output from Group One can be seen in Fig. \ref{fig:group1visual}.

To substantiate the reliability and generalizability of the proposed segmentation approach, we conducted a validation study using an independent dataset of five porcine eyes (Group Three), for which the algorithms were not previously trained and for which the statistical approach was not tested. The validation is designed to mirror how the algorithm and network would generalize on unseen patients. The validation commenced with applying Gamma distribution fitting to patches of the B-scans of the new porcine eyes, creating two patch-wise 2d matrices, one for parameter alpha and the other one for parameter beta. Subsequently, the tool, ILM, and RPE parameters were isolated based on the Random Forest algorithm previously trained on Group One and Group Two. The configurations A, B, C, and D were then tested with dataset Group Three for the networks trained on Group One and Group Two. As can be seen in Table. \ref{tab:porcineye_result} the results with configuration with B-scans at input (A) had the lowest mean DICE score of 0.10 and the biggest mean Hausdorff distance of 351 while the configuration consisting of the Alpha and Beta parameter (D) matrices performed the best with a Mean DICE score of 0.90 and the lowest mean Hausdorff distance of 141. The configuration consisting of weak binary maps as input (B) performed better than the configuration consisting of  B-scans, Alpha matrix, and Beta matrix. With a mean DICE score of 0.87 compared to 0.75.

\begin{table}[H]
\centering
\begin{tabular}{lcccc}
\hline
\multicolumn{5}{c}{\textbf{Group Three results}} \\ \hline

 Training Configuration &\textbf{A} & \textbf{B } & \textbf{C}   & \textbf{D}           \\ \hline
 \textbf{DICE} & &  &   &            \\ \hline
Mean        & 0,10      & 0,87             & 0,75       & 0,90        \\ 
Min          & 0,05     & 0,63             & 0,20       & 0,85        \\ 
Max          & 0,99     & 0,99             & 0,99       & 0,99        \\ \hline
 \textbf{Hausdorff} & &  &   &            \\ \hline
Mean        & 351      & 163             & 232       & 141        \\ 
Min          & 221     & 24             & 33       & 17       \\ 
Max          & 475     & 474             & 475      & 474        \\ \hline

\end{tabular}
\caption{DICE score and Hausdorff Distance of Group Three}
\label{tab:porcineye_result}
\end{table}

\begin{table}[h]
\centering
\begin{tabular}{lcccc}
\hline
\multicolumn{5}{c}{\textbf{Group One and Group Two results}} \\ \hline

 Training Configuration &\textbf{A} & \textbf{B } & \textbf{C}   & \textbf{D}           \\ \hline
 \textbf{DICE} & &  &   &            \\ \hline
Mean        & 0,96      & 0,95             & 0,96       & 0,97\\
Min          & 0,93     & 0,91             & 0,94       & 0,95\\
Max          & 0,99     & 0,99             & 0,99       & 0,99        \\ \hline
 \textbf{Hausdorff} & &  &   &            \\ \hline
Mean        & 45      & 33             & 20       & 29\\        
Min          & 4     & 4             & 4       & 0\\       
Max          & 330     & 350             & 350      & 351        \\ \hline

\end{tabular}
\caption{DICE score and Hausdorff distance of Group One and Group Two}
\label{tab:group1and2_result}
\end{table}

\begin{figure}[h]
\centering
    % Three
    \includegraphics[width=\columnwidth]{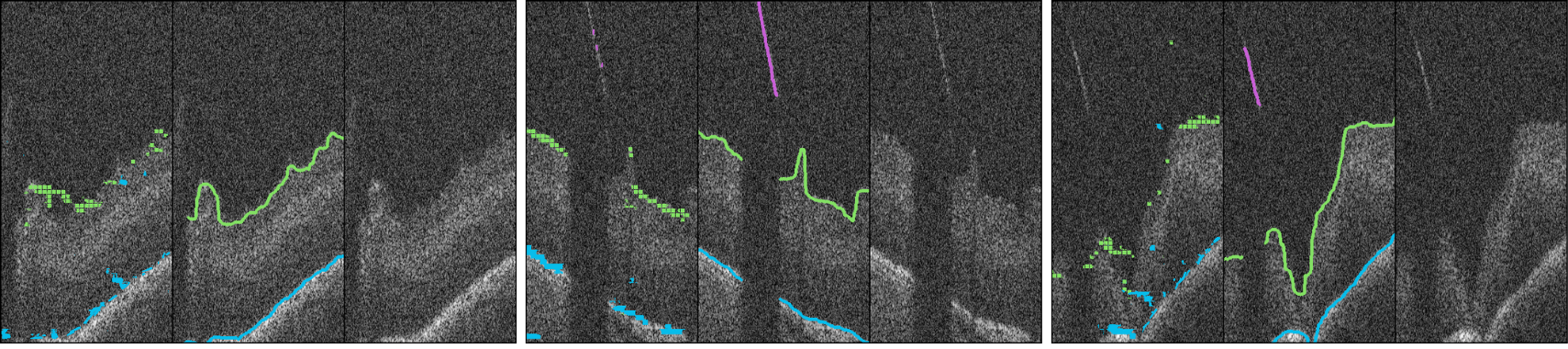}
    \includegraphics[width=\columnwidth]{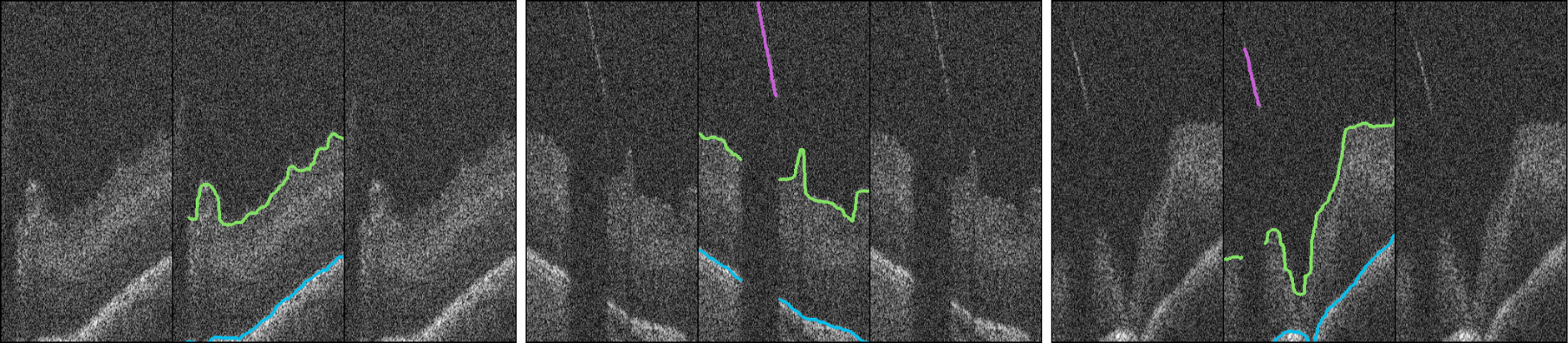}
    \caption{Three examples of Group Three images with training configuration D (top) and training configuration A (bottom). In each example, the left image is the output, the middle image is the ground truth, and the right image is the raw B-scan.}
    \label{fig:GR3_visual}
\end{figure}

The detailed analysis across various network configurations and metrics emphasizes Configuration D, which utilizes Gamma parameters exclusively as the most effective. It outperforms others with the highest accuracy, Dice metric, and lowest Hausdorff distance, as detailed in Table. \ref{tab:porcineye_result}. Interestingly, while Configuration D led in performance, other configurations, including binary maps only (B) and the combination of B-scans and Gamma parameters (C), also demonstrated valuable segmentation capabilities. The visualization of the resulting segmentations for Group Three can be seen in Fig. \ref{fig:GR3_visual}

This comprehensive analysis showcases our segmentation method's nuanced and adaptable nature, especially when using Gamma parameters and weak labels. The method's demonstrated precision and adaptability promise advancements, paving the way for its clinical integration.

\section{Discussion}
\label{sec:discussion}
In the context of the study, the comparative analysis of the distribution fitness for B-scan Groups One and Two yields significant insights. Our findings indicate that Gamma and Burr distributions demonstrated the best overall fit of distribution to scatter. Rayleigh, Normal, and Gamma distributions also marked an elegant separation between the parameter values based on class belonging, simplifying the segmentation process with easy value isolation to create binary maps. The analysis, backed by prior research on biological tissues, reveals that the Gamma distribution provides a reasonable fit with acceptable variability and good parameter value separation, which is a commendable balance between accuracy and reduced manual labor. This balance is pivotal as it directly impacts the practicality of iOCT applications in surgical settings, where both low manual labor and precision are important. Hence, steps two and three of the pipeline shown in Fig. \ref{fig:pipeline} were processed with the Gamma Distribution.

The testing of the pipeline with the data from Group Three illustrated the segmentation performance across retinal layers and tools in previously unanalyzed data sets. Notably, using Gamma-fitting-derived binary maps (B) as an input enables our network to refine segmentation layers with high accuracy. The resilience of the segmentation network is further demonstrated through its adaptability to new data without necessitating retraining using Gamma-parameter matrices (D) as input. This adaptability underscores the robust combination of statistical speckle measures and biological insight in segmenting unfamiliar data sets. The method showcases high fidelity and overcoming challenges posed by variability and noise inherent in in-vivo imaging. 

The visual results in Fig \ref{fig:GR3_visual} and the metric results in Table \ref{tab:porcineye_result} show that B-scans alone (configuration A) as input is not enough for the segmentation network to be robust to the variability of the retinal structures in different eyes and time points, even for the same OCT device. This could be because the network is basing the estimation on shape and pixel intensity, which can vary depending on the porcine eye and device settings, resulting in data outside of the training distribution. On the other hand, using the Gamma parameters (D) as input results in a network invariant to shape and light variations since it bases the estimation on the scattering properties of different mediums, independent of the shape. Proving the opportunities of using Gamma distribution to segment retinal layers of unseen samples.

One notable limitation of our current approach is the differentiation of surgical tools within the OCT scans As shown in Fig. \ref{fig:GR3_visual}, the thin semitransparent needle profile in the random test dataset is not prominent enough for the Gamma distribution to differentiate. Tools made of different materials, such as metal and glass, present distinct challenges for segmentation methods. With their opaque nature, metallic tools generally produce strong reflections that are identifiable in OCT images. Conversely, tools made from semitransparent materials pose a significant challenge for segmentation algorithms due to their less prominent signal. Moreover, the process of optimizing parameters for new surgical tools is time-consuming and labor-intensive. Each new tool type introduced into the surgical field requires an adjustment of the statistical parameters to ensure accurate representation within the OCT images. To mitigate the challenges associated with manual parameter optimization for new tools, a prospective direction for the research is automating the parameter-finding process. Developing an algorithmic solution that can dynamically adjust parameters in response to introducing new tool types would substantially reduce the time and expertise required for effective segmentation. This automation would enable a more efficient adaptation to diverse surgical environments and tool sets, thereby enhancing the clinical applicability of our OCT imaging methodology.

In conclusion, the analysis and validation provide strong evidence for the practicality and reliability of utilizing different statistical scattering properties in order to segment retinal layers in porcine eyes. These findings have significant implications for the future of automated segmentation in clinical ophthalmology, potentially enhancing the speed and
precision of procedures.

\bibliographystyle{plain} % We choose the "plain" reference style
\bibliography{refs} % Entries are in the refs.bib file

\end{document}